\documentclass[a4paper, twocolumn]{article}
\usepackage{epsfig}
\usepackage[dvips]{color}
\usepackage{amsmath,amssymb}

\newcommand{\be}{\begin{equation}}
\newcommand{\ee}{\end{equation}}
\newcommand{\bea}{\begin{eqnarray}}
\newcommand{\eea}{\end{eqnarray}}
\newcommand{\ba}{\begin{array}}
\newcommand{\ea}{\end{array}}
\def\bra{\langle}
\def\ket{\rangle}

\begin{document}
\title{Resonant Activation in Asymmetric Potentials}

\author{Alessandro Fiasconaro\footnote{E-mail address: afiasconaro@gip.dft.unipa.it}\\
{\em Mark Kac Complex Systems Research Center and
Marian~Smoluchowski Institute of Physics} \\ {\em Jagellonian
University, Reymonta 4, 30-059 Krak\'ow, Poland}, \\and  \\
{\em Dipartimento di Fisica e Tecnologie
Relative\footnote{Group of Interdisciplinary Physics,
http://gip.dft.unipa.it} and CNISM, Universit\`a di
Palermo} \\ {\em Viale delle Scienze, I-90128 Palermo, Italy}
}
%
%
%
\maketitle

\begin{abstract}
The resonant activation effect (RA) has been well studied in
different ways during the last two decades. It consists in the
presence of a minimum in the mean time spent by a Brownian
particle to exit from a potential well in the presence of a
fluctuating external force, as a function of the mean frequency
(or the correlation time) of the latter. This work studies the
role played by the asymmetry of a piecewise linear potential in
the RA effect, and, in general, the behavior of the mean first
passage time and the mean velocity of the particle crossing
through the potential barrier. A strong dependence on the
asymmetry of the potential has been found which can be put in
relationship with the current in the ratchet whose the potential
here used is an elementary module. In this case a current reversal
as a function of the frequency of the switching potential occurs.
Comparison of the calculations with the Doering-Gadua model have
been performed, as well as comparison with smooth symmetrical
potentials, by checking for the robustness of the resonant
correlation time. The calculations have been done by solving
numerically the Langevin equation in the presence of an
uncorrelated Gaussian noise. The resonant mean first passage times
show an unexpected behavior as a function of the thermal noise
intensity. The related curves present for the different symmetries
an unexpected inversion of their relative behavior beyond a
certain threshold value of the noise. This means that the current
reversal can only occur for weak noise intensities, lower than
that threshold value.
 \vskip 0.2 cm
 \noindent
  Pacs: 05.40.-a, 05.45.-a
 \vskip 0.5 cm
\end{abstract}
In the recent past years various theoretical works have been
produced around the concept of Resonant Activation (RA)
~\cite{doe,bie,hanggi-ra,rei,iwa,boguna,bdka}, which consists in
the presence of a minimum of the mean escape time from a potential
well of a Brownian particle when the system is subjected to a
randomly switching force, as a function of the mean switching
quantity. The RA effect has been also detected experimentally
\cite{ms,schmitt,sun} and it is in principle involved in a wide
branches of science from physics to biology. The occurrence of the
RA together with other stochastic effects such as noise enhanced
stability (NES) \cite{nes,sun,alenes} and stochastic resonance
(SR) \cite{gammaitoni} have been also investigated
\cite{schmitt,alePRE2006}. The article by Doering \& Gadua
\cite{doe} is considered as one of the most introductory work to
the resonant activation phenomenon. They introduced a switching
piecewise linear potential in a range $[0,L]$ with fixed minima in
$x=0$ and $x=L$, and fluctuating amplitude of the maximum. They
report the mean escape time for a Brownian particle in the case of
fluctuations of the maximum of the potential between $V_0$ and $0$
(flat potential), and also between $V_0$ and $-V_0$ (well instead
of barrier), showing the presence of the RA effect. A slightly
different choice was made by Bier \& Astumiam \cite{bie} who used
a potential fluctuating between $V_0-a$ and $V_0+a$ with $a<V_0$,
maintaining so the presence of the barrier in all the dynamics.
Both the methods show the resonant activation effect, and have
been analytically evaluated in some approximation \cite{boguna}.
Both the choices have in common that the potential is symmetrical
in shape and maintains the same value at the two extrema in all
the dynamics ($V(0) = V(L)$).

Aim of this work is to focus on the role played by the asymmetry
of the potential on the resonant activation effect using a simple
piecewise linear potential .

Many papers with both experimental and analytical investigation
concerning the role played by the asymmetry of the potential in
stochastic effects have been published during the last years.
However, the investigations have been mainly devoted to the
effects on the Stochastic resonance phenomenon, that is the
presence of a noise induced regular oscillations in a system,
which is revealed by means of a maximum in the signal to noise
ratio of the output \cite{dykmann,wio,li,stocks,ning}, while the
relation between the RA and the shape of the potential has been
previously performed using a single slope linear potential
\cite{dybiec}.

Comparison with the Bier-Astumian model have been performed, as
well as comparison with smooth symmetrical polynomial potentials.
The results obtained have been extended to the most elementary
ratchet potential, giving explanation of the current reversal
there found as function of the correlation time of the external
force.

The fluctuating potential $V_{\pm}(x,t)$ is here given as the sum
of a static potential $V(x)$ (with, again, $V(0)=V(L)$) plus an
additional time-dependent $U(x,t)$ giving the two configuration
'up' and 'down' between which $V_{\pm}(x,t)$ takes its values. The
additional potential $U(x,t)$ has not to be necessarily a
stochastic process to give rise to RA \cite{rei,boguna}. It can be
a smooth, continuous potential like a cosine or, instead, a
stochastic potential related to a dichotomous force exponentially
correlated in time. This last form is widely used in literature
and we use it in this work. The related Langevin equation is:
 \be
   \dot{x} = -V'(x) + \eta(t) + \xi(t)
   \label{langevin}
 \ee
where $\xi(t)$ is the Gaussian white noise, with zero mean and
correlation function $\bra\xi(t)\xi(t')\ket=2D\delta(t-t')$. The
intensity $D$ is related to thermal bath and damping coefficient
$\gamma$ (here $\gamma=1$) by means of the relation $D=\gamma k_B
T$. The random force $\eta(t)$ represents a dichotomous stochastic
process, the random telegraph noise (RTN), taking the two values
$\{-a,a \}$ with an exponential correlation function $\bra \eta(t)
\eta(t')\ket = (Q/\tau) e^{-|t-t'|/\tau}$, where the intensity $Q
= a^2 \tau$ and $\tau$ is the correlation time of the process.
 \begin{figure}[htbp]
 \begin{center}
 \vskip -1.2 cm
  \includegraphics[angle=0, width=8.5cm]{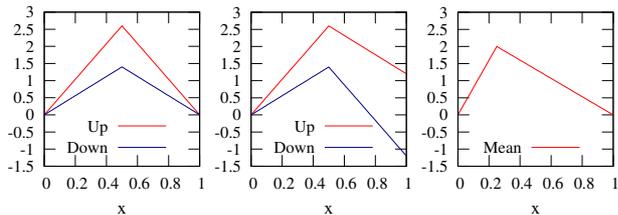}
 \vskip -0.7 cm
\caption{Piecewise linear potential used in the calculation. The
position $x_0=0$ represents the starting point of the simulations
and there is also put a reflecting boundary. With respect to
Bier-Doering choice (left) the right extremum of the potential is
not fixed (center) and the whole potential flips randomly between
the two shapes $\{V_+, V_- \}$ with a correlation time $\tau$. The
right draw shows an example of asymmetric piecewise linear static
potential. The parameter $k$ represents the position of the
maximum $x_m$ with respect to the position of the maximum in the
symmetric case $x_s$ (here $x_s=0.5$, $x_m=0.25$, $k=-0.25$)}
  \label{pot}
 \end{center}%
\end{figure}
The potential $V_{\pm}(x,t)$ is then defined as:
 \be
   V_{\pm}(x,t) = V(x) + U(x,t) = V(x) -x \eta(t).
   \label{Potential}
 \ee
 with, explicitly,
 \be
   V(x) = \left\{
   \begin{array} {lr}
    h \frac{ x}{x_m}     &    0 \le x_m  \\
    h \frac{L-x}{L-x_m} &  x \ge x_m
   \end{array} \right.
   \label{Potential-s}
 \ee
Here $L=1$, $h=2$, $x_m = L/2 + k$, and $k$ represents the
asymmetry parameter, defined as the distance of the position of
the maximum of the potential $x_m$ from the position of the
symmetrical maximum $x_s$. In Fig. \ref{pot} (right) we see an
example of the static potential with the asymmetry parameter
$k=-0.25$.

The difference between the choice of the fluctuating potential
here used (Eqs. \ref{Potential} and \ref{Potential-s}) and that
one by Bier-Astumian is visible in Fig. \ref{pot}, where the
potentials are drawn in the two cases. Here, with respect to the
'mean' static potential, the additional fluctuations $-x\eta(t)$
give only two values ('up' and 'down' in fact) in the flipping,
being the force $\eta(t)$ uniform overall the $x$-range of the
potential. In the Doering \& Gadua model (as in the Bier-Astumian)
two values of the force for each potential slope have to be
considered to hold the minimum on the right at the same level
($V(L)=const$).

The potential $V_{\pm}(x,t)$ here defined can be considered as a
base modulus of the piecewise linear ratchet subjected to RTN
widely used in literature \cite{Magnasco,Czernik,Kula}.
\begin{figure}[htbp]
 \begin{center}
  \includegraphics[angle=-90, width=8.5cm]{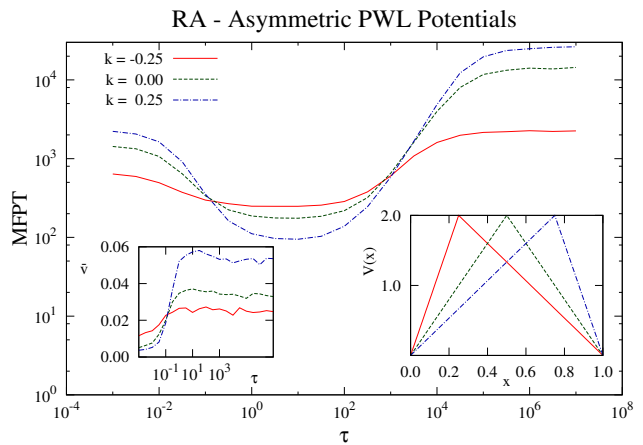}
\caption{MFPT showing the resonant activation effect for three
values of the asymmetry parameter $k$ of the piecewise linear
potential ($k=-0.25$, $k=0$, such as symmetric potential,
$k=0.25$). The white noise intensity is here $D=0.18$. The
intensity of the dichotomous force is $a=1.2$. The bottom/left
inset show the mean velocity of the Brownian particle for the same
asymmetries, as a function of the correlation time $\tau$.}
 \label{PWL-RA}
 \end{center}%
\end{figure}
The equation (\ref{langevin}) has been solved numerically by using
$dt=10^{-3}$ and the averages have been performed over a sampling
of $N=20,000$ realizations. In the $i$-th realization the particle
is put in the starting position $x_0 = 0$ and the time $t_i$ to
cross the position $x=L$ is computed. A reflecting boundary is put
in the left extremum of the potential while an absorbing boundary
is present at the right extremum. The ensemble average of the
$t_i$ gives the Mean First Passage Time (MFPT), which presents,
for all the cases here studied, the evidence of the RA effect,
i.e. a well drawn minimum as a function of the correlation time
$\tau$. In fact, as well as the symmetric case, the MFPT obtained
with the asymmetrical potentials show a resonant effect which is
drawn in Fig. \ref{PWL-RA}. We notice that for the three values of
the asymmetry parameter $k$, we find quite the same value of the
resonant correlation time $\tau_R \sim 10$, but different values
of the corresponding resonant MFPTs ($T_Rs$), which decrease by
increasing the asymmetry parameter $k$.

We note that the resonant region shows an inversion in the
behavior of the MFPT curves for the three potentials to both the
low and high correlation times with respect to the intermediate
one. In fact for $\tau$ lower than $\tau_{C_L} \approx 10^{-1}$
the curves show a MFPT higher for positive asymmetry ($k=0.25$)
and lower for negative asymmetry ($k=-0.25$) and the same
qualitative behavior is visible in the long correlation time
region $\tau$ higher than $\tau_{C_R} \approx 10^{3}$. In the
intermediate region $\tau \in [\tau_{C_L},\tau_{C_R}]$, where we
also find the resonant values, the situation is inverted: the
highest $T_R$ value corresponds to the negative asymmetry
parameter and the lowest $T_R$ to the positive one.

\begin{figure}[htbp]
 \begin{center}
  \includegraphics[angle=-90, width=8.5cm]{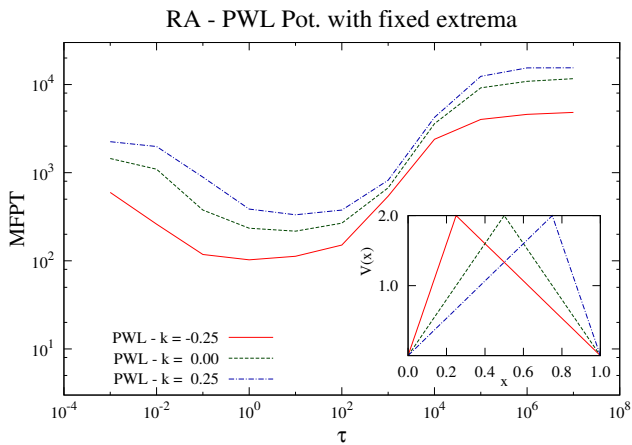}
\caption{MFPT showing the resonant activation effect for different
asymmetrical piecewise linear potential with fixed extrema and
equal potential excursion in the three cases. Differently from the
results in Fig.~\ref{PWL-RA} we don't find any cross between the
three curves and also the resonant correlation time $\tau_R$
changes for the different asymmetries. The potential increase is
here $a=0.6$, as the one of the symmetric case in
Fig.~\ref{PWL-RA}.}
 \label{PWL-RA-FIX}
 \end{center}%
\end{figure}

\begin{figure}[htbp]
 \begin{center}
  \includegraphics[angle=-90, width=8.5cm]{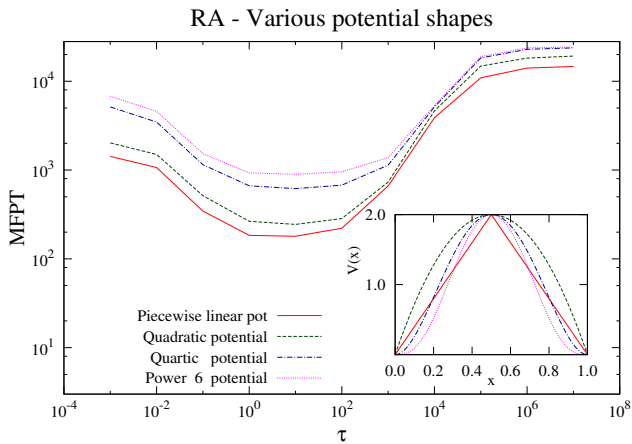}
\caption{MFPT showing the RA effect for different shape of
symmetrical potential. We note than for $\tau \lesssim \tau_R$ the
logarithmic distances between two MFPTs is approximatively a
constant. This evidence suggests that in that region an
exponential from factor should be inserted in the analytical
expression of the mean first passage time to take into account the
shape of different potentials. The parameters are the same than in
Fig.\ref{PWL-RA}. }
 \label{shapes}
 \end{center}%
\end{figure}

On the other hand, calculation performed with the Bier-Astumian
and Doering \& Gadua model, that is using fixed extrema of the
same asymmetric piecewise linear potentials, which fluctuates
between the same highs ($a=0.6$) in all the asymmetries, give
strongly different curves, as visible in Fig.~\ref{PWL-RA-FIX},
where we can see even a strong displacement of the resonant
correlation time by changing the asymmetry, but no crosses are
present between the MFPT curves.
The inversion of the MFPTs curves behavior, and consequently the
presence of the two crosses at approximatively $\tau_{C_L}$ and
$\tau_{C_R}$, is so uniquely present in MFPTs calculated for
asymmetrical potentials using uniform fluctuating force overall
the range $[0,L]$, and it does not appear neither in the
symmetrical ones with different shapes (See Fig.~\ref{shapes}),
nor in the asymmetrical ones with fixed extrema and fluctuating
barriers (Fig.~\ref{PWL-RA-FIX}). In other words, the comparison
between the results plotted in Figs.~\ref{PWL-RA},
\ref{PWL-RA-FIX} and \ref{shapes} put in evidence that the cross
features of the MFPT curves occurs not merely because of the
asymmetry of the potentials, but, instead, because of the presence
of the asymmetry \textit{together} with the uniformity in space of
the fluctuating external force $\eta(t)$ added to the system.

The main relevant feature in adding a uniform force in the range
of the constant potential, lies in the fact that in this case the
barrier high of the fluctuating potential takes different values
for different positions of the maximum, i.e. as a function of the
asymmetry parameter $k$. In fact the resonant MFPT values $T_R$s
depend mainly by the lower value taken by the potential ($V_-$),
being proportional to $(1/V_-^2) e^{V_-/D}$ \cite{boguna}, and
this value becomes lower and lower, by increasing the value of
$k$s. This means that at a first sight we can expect that the
$T_R$ values take a lower value for the positive asymmetry than
for the negative ones. However, as we can see below in the text,
this expectation holds only up to a certain threshold value of
noise intensity ($D_T$) and the inverse behavior occurs for higher
values ($D > D_T$).

The model here investigated presents interesting features in the
MFPT: first of all it has a value of the resonant mean period
$\tau_R$ not too strongly dependent on the asymmetry parameter
$k$; then, it presents two period intervals, close to $\tau_{C_L}
\approx 10^{-1} $, and close to $\tau_{C_R}\approx 10^{3}$ having
approximatively the same MFPT for all the $k$-parameters.

\begin{figure}[htbp]
 \begin{center}
  \includegraphics[angle=-90, width=8.5cm]{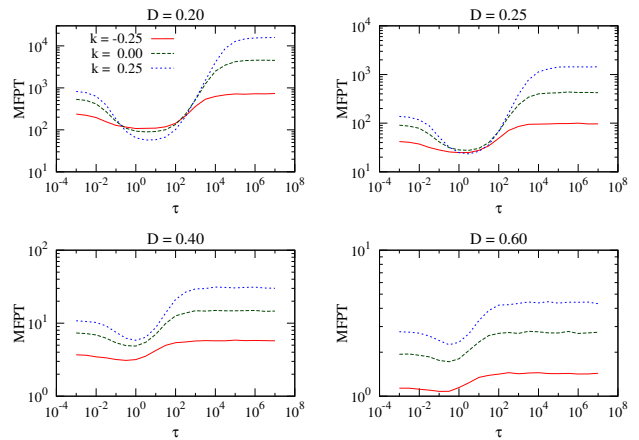}
\caption{Resonant activation evolution for various noise
intensities. We observe the disappearing of the crossings in the
MFPT curves and the shift of the minima by increasing the noise
intensity $D$. These behaviors are shown in details Fig.
\ref{omega-T-vs-D}, where the resonant frequency  and resonant
times are plotted as a function of $D$.}
 \label{RA-D}
 \end{center}%
\end{figure}

\begin{figure}[htbp]
 \begin{center}
  \includegraphics[angle=-90, width=8.5cm]{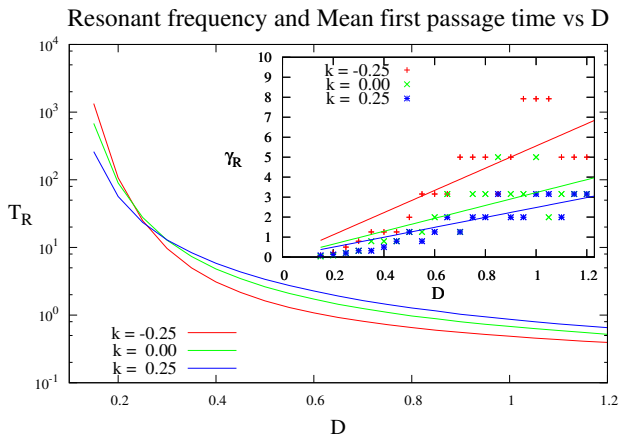}
\caption{Resonant mean passage times $T_R$s as a function of the
noise intensity $D$ for the three asymmetry values investigated.
In the inset the corresponding resonant frequencies $\gamma_R$s.
The crossing value $D_T \approx 0.27 $ represents the threshold of
thermal noise discriminating if the two crossings in MFPT of
Fig.~\ref{PWL-RA} are present ($D<D_T$) or absent ($D>D_T$).}
 \label{omega-T-vs-D}
 \end{center}%
\end{figure}

However, the crossing features of the MFPT as a function of the
mean driving frequency doesn't occur for any value of the thermal
noise intensity $D$. A set of calculation to check this kind of
robustness has been performed and the related results are shown in
Fig.~\ref{RA-D}, where the RA is plotted for different values of
the noise. We can see that by increasing the thermal noise
intensity $D$, the crosses between the curves are maintained up to
a threshold value that we can call $D_{T}$. For higher noise
intensities no crosses appear in the curves. Further, a shift of
the resonant mean switching time as a function of noise is visible
and a lowering of the related $T_R$, which demonstrate that when
the noise is increased, an higher frequency switching is necessary
to reach the resonance, and this resonance occurs at a lower mean
escape time. The increase of the noise intensity has in this sense
the effect to speed up all the escape features from the well. The
results shown in Fig.~\ref{RA-D}, can be better observed in
Fig.~\ref{omega-T-vs-D}, where the mean resonant frequencies
$\gamma_R = 1 / (2 \tau_R)$ and the mean resonant escape times
$T_R$s have been plotted as a function of the noise intensity for
the three symmetry values. We can see there that going beyond the
threshold noise value $D_T$, the three $T_R$ curves invert their
relative position. This noise threshold corresponds to the
presence (for $D<D_T$) or the absence (for $D>D_T$) of the two
crossings of the MFPT curves visible in Fig.~\ref{PWL-RA} and
Fig.~\ref{RA-D}. However we notice that $D_T$ is not unique for
all the asymmetries, being the crossing values slightly different
for each couple of the three curves. We can see that the resonant
frequency (inset of Fig. \ref{omega-T-vs-D}) has a slightly
different dependence on the thermal noise intensity $D$ for the
different asymmetries. In the range of $D$ investigated, the three
curves can be easily approximated by a straight line, even if the
real dependence is in general more complicated (see
\cite{boguna}); the slope of this line is higher for the negative
asymmetry than for the positive one.

The presence of a resonant behavior, as well as the cross value at
$\tau_{C_L}$, is also found in the plot of the mean velocity of
the Brownian particle. Left inset of Fig.\ref{PWL-RA} shows, in
fact, this measure as a function of the correlation time of the
fluctuating dichotomous force, calculated as $\bar{v}=N^{-1}
\sum_{i=1}^N L/t_i $. For all the asymmetry parameters, we see the
presence, before the saturating behavior, of a weak maximum which
corresponds to the resonant correlation time $\tau_R$. We can also
see that for low values of the correlation times ($\tau <
\tau_{C_L}$) the mean velocity is higher for negative asymmetry
and lower for positive ones, while for ($\tau > \tau_{C_L}$) is
the inverse. This feature gives rise to a reversal current in the
ratchet, as predicted in other works
\cite{ReiEst,dykmann_cr,luchinsky} and whose occurrence has been
also demonstrated experimentally \cite{gommers}. In fact the
difference between the mean velocities of the positive asymmetry
and the negative one change sign at the $\tau_{C_L}$ value. In an
asymmetrical ratchet this difference represents a net velocity
flux, provided that the absence of any reflecting boundary in that
case gives rise to changes in the values of the velocity. Both the
presences of a maximum for $\tau \approx \tau_R$ and the cross at
$\tau \approx \tau_{C_L}$ are in total agreement with the behavior
of the MFPT. This agreement fails, instead, for values of the
correlation times higher than $\tau_R$. While the MFPT curves
increase in a different way and joint together at the second
cross, the velocities decrease only a few, reaching a saturation
value. This is because for high values of the correlation times,
the particle tends to cross the potential barrier when it is in
its lower high, so acquiring a relatively high speed because of
the low travelling time. When the potential is in the high level,
the particle takes a longer time to cross and so the contribution
to the mean velocity becomes very low and relatively negligible.
This means that, for high correlation times, $\bar{v}$ maintains a
relatively high value which doesn't change so strongly as the
MFPT.
\begin{figure}[htbp]
 \begin{center}
  \includegraphics[angle=-90, width=8.5cm]{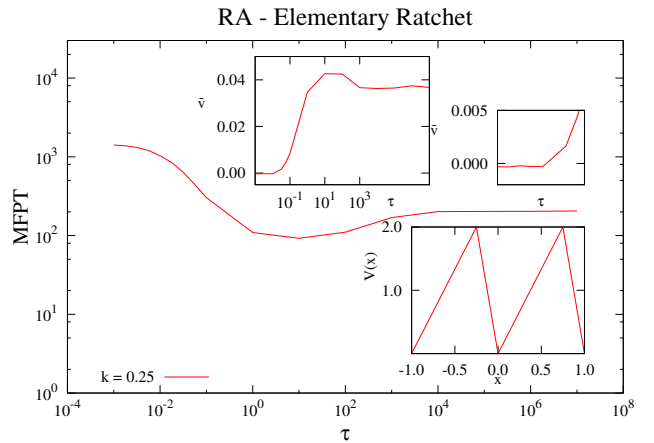}
\caption{MFPT showing the resonant activation effect for the
simplest ratchet potential, composed by only two the single
barrier. The MFPT is here the mean time taken by the Brownian
particle starting at $x=0$ to reach the position $x=1$ \textit{or}
$x=-1$. The mean velocity of the particle is plotted in the
upper/left inset. The mean velocity has again a maximum at the
same resonant value $\tau_R$. In the very low correlation time
region, the system has a weak negative velocity (right/top inset),
as a consequence of the different behavior and the cross of the
mean velocities shown in the inset of Fig.~\ref{PWL-RA}.}
 \label{ratchet}
 \end{center}%
\end{figure}

The results found above for the single barrier potentials,
mirrors, of course, to the ratchet potential having the same
asymmetric profile as elementary module. In this respect a set of
calculations has been performed with the aim to join together the
results of the single barrier described above with the simplest
ratchet case, such as a ratchet with two barriers only. Fig.
\ref{ratchet} shows the results in such a case and the
bottom/right inset shows the corresponding elementary ratchet. The
system consists of two asymmetric barriers without the presence of
any reflecting boundary. The MFPT presents again a resonant
correlation value $\tau_R$ which is the same for the single
barrier case, as we can expect. In this system the MFPT is the
mean time spent by the Brownian particle starting at $x=0$ to
reach the position $x=1$ \textit{or} $x=-1$, indifferently. The
particle, of course will follow the easiest path, and the MFPT
represents the minimum time of the two single barrier case seen
above. This also means that the curve is lowered and the RA effect
less pronounced. The mean velocity, plotted in the upper-left
inset of the Fig.~\ref{ratchet}, shows again a maximum at the same
resonant value $\tau_R$. For very low correlation time the mean
velocity has a weak negative velocity (right/top inset in
Fig.\ref{ratchet}). This means that a current reversal appears at
a certain correlation time $\tau_{rev}$. This features follows
from the different behavior of the mean velocity in the two
specular asymmetric single barrier potentials seen above (inset of
Fig.\ref{PWL-RA}), where the presence of the cross value
$\tau_{C_L}$ indicates a current reversal as a function of $\tau$.
The difference in value between $\tau_{rev}$ and $\tau_{C_L}$, as
well as the difference in the absolute value of the mean velocity
of the Brownian particle, have to be imputed to the presence of
the reflecting boundary in the single barrier case which change
the traveling times of the particle and, so, the related mean
velocities.

As a last remark concerning the relationship between the resonant
activation effect and the shape of the potential, some
calculations have been performed using symmetrical smooth
potentials. The static potentials used have the form:
 \be
   V_{2N}(x) = h \; 2^{2N} \frac{x^N}{L^N} \left( 1-
   \frac{x}{L} \right )^N
 \ee
where in our calculation $h=2$. The values used are: $N=1,2,3$,
such as parabolic, quartic and $6^{th}$ power potentials. As we
can see in Fig. \ref{shapes}, the resonant mean time is quite the
same for all the cases, again confirming that $\tau_R$ is a robust
value in the model investigated. Another remarkable and well
visible feature is that the four curves of the MFPT differ each
other of a constant quantity, at least for $\tau \lesssim \tau_R$.
This means that, in that region, their logarithmic distance is
constant and so an exponential form factor has to be taken into
account in order to estimate the MFPT for each potential shape.

Summarizing the results, the shapes of the potential (both
symmetrical and asymmetrical ones) play a very important role in
the evaluation of the RA effect and MFPT behaviors. With a
spatially uniform random telegraph force, the resonant correlation
time $\tau_R$ appears to be a robust value independently on that
shape, while this latter acts always in a strong way by modifying
the resonant values of the mean first passage times $T_Rs$. In the
context of uniform forces, the asymmetry of the potential is then
responsible for the crosses of the MFPT curves in a certain range
of low thermal noise intensities, giving an explanation for the
appearance of the current reversal as a function of the
correlation time of the fluctuating force in ratchet potentials.
These crosses, and, consequently the current reversal in ratchet,
are only present at weak noise intensity, as indicated by the
presence of an upper noise intensity threshold $D_T$.

\vspace{0.5cm} This work has been supported by the Marie Curie TOK
grants under the COCOS project (6th EU Framework Programme,
contract No: 52/MTKD-CT-2004-517186).

\end{document}